\documentclass[runningheads]{svmult}
\usepackage{makeidx}   
\usepackage{graphicx}  
\usepackage{subeqnar}  
\usepackage{multicol}  
\usepackage{physprbb}  
\makeindex             

\begin{document}
\title*{Distances 
to Local Group Galaxies}
\toctitle{Distances to
Local Group Galaxies}
%
%
\titlerunning{Distances to Local Group Galaxies}
%
\author{Alistair R. Walker}
\authorrunning{Alistair R. Walker}
%
%
\institute{Cerro Tololo Inter-American Observatory, NOAO, Casilla 603, la Serena, Chile}

\maketitle              

\begin{abstract}

Distances to galaxies in the Local Group are reviewed.  In particular, the distance
to the Large Magellanic Cloud is found to be $(m-M)_0 = 18.52 \pm 0.10$, corresponding to
$50,600 \pm 2,400$ pc.  The importance of M31 as an analog of the galaxies observed at
greater distances is stressed, while the variety of star formation and chemical enrichment
histories displayed by Local Group galaxies allows critical evaluation of the calibrations
of the various distance indicators in a variety of environments.

\end{abstract}

\section{Introduction}

The Local Group (hereafter LG) of galaxies has been comprehensively described in the monograph
by Sidney van den Berg \cite{vdb1}, with update in \cite{vdb2}.  The zero-velocity surface has
radius of a little more than 1 Mpc, therefore the small sub-group of galaxies
consisting of NGC 3109, Antlia, Sextans A and Sextans B lie outside the 
the LG by this definition, as do galaxies in the direction of the nearby
Sculptor and IC342/Maffei groups.  Thus the LG consists of two
large spirals (the Galaxy and M31) each with their entourage of 11 and 
10 smaller galaxies respectively, the dwarf spiral M33, and 13 other galaxies classified as
either irregular or spherical.  We have here included NGC 147 and NGC 185 as members of the
M31 sub-group \cite{vdb4}, whether they are actually bound to M31 is not proven.
Similarly, Leo I and Leo II are classified as satellites of our Galaxy, however \cite{vdb1}
has pointed out that the mass of our Galaxy becomes uncomfortably large if they are
indeed bound.
Of these 36 galaxies, 23 are classified \cite{vdb1},
\cite{vdb2} as being dwarf galaxies with $M_V < -14.0$. There are no giant ellipticals, 
the nearest being some 7 Mpc distant in the Leo I group, nor is there anything so
exotic as NGC 5128 (Cen A) at 4 Mpc distance in the Centaurus group.  However
there are some interesting `one-off's'; M32 is a dwarf elliptical, and IC 10 is
an irregular galaxy presently undergoing very active star formation (starburst galaxy).
The LG as defined above is listed in Tables 1-4.  Columns 1-3 give the galaxy
name, type and approximate absolute magnitude \cite{vdb1}, \cite{vdb2}, while column 4 gives
an indication of the population mix, which is a guide to the types of distance
indicator present.  The star formation history of local group dwarf galaxies 
is remarkably diverse, and the true situation is much more complex than this simple
guide, which has divisions of young (less than $\sim 1$ Gyr), intermediate (1-7 Gyr), and
old (7-12 Gyr).  Throughout, old populations appear to be ubiquitous, even though their
fractional contribution to the total light can be very small, and it is not clear
whether the formation times are coincidental, or spread over a few Gyr
\cite{greb}.

The LG is contained in what is termed the `Local Volume', a sphere with 
radius approximately 10 Mpc, thus a a factor 1000 times the volume of the LG.
A systematic census of galaxies likely to lie in this volume \cite{krannk}, those with
$V_{LG} < 500 $ km/s, listed 179 members, this number has been
doubled by more recent work \cite{hucht}. These galaxies are clustered in
rather ill-defined groups, with substantial volumes (e.g. the `Local Void')
free or almost free of galaxies.  The closest groups have zero-velocity
surfaces that are close to that for the LG, for instance the Sculptor
group appears to be very elongated and viewed almost end-on, the nearest
members such as NGC 55  are less than 2 Mpc from the LG barycenter.  The Centaurus
group, which is estimated to be about seven times as massive as the LG
\cite{vdb3} has zero-velocity surface only $\sim 2$ Mpc from the LG barycenter.
The large numbers of dwarf galaxies recently found in both groups 
\cite{jerj} appear more spatially dispersed than do the more massive galaxies,
this is also true for the LG.

As far as we know, LG galaxies are typical of the `mean' population of
galaxies, thus a detailed study should allow deductions to be made 
concerning the general properties of galaxies, in particular their formation and 
subsequent evolution,
throughout the Universe.  The common dwarf spheroidal (dSph) galaxies are
the best places to test the small scale predictions of hierarchical galaxy
formation models, and the nature and distribution of dark matter \cite{gil}.
Indeed, the favored cold dark matter (CDM) formation theory predicts a factor
10 more dSph galaxies in the LG than are known, however it is estimated \cite{mat}
that we have found more than half of them. Recent successful \cite{whit} and
on-going \cite{will} searches are helping to refine the total numbers of LG
dSph's, but we have found all the
higher surface-brightness members unless they are hidden directly behind the
galactic plane.
  
The latest generation of large telescopes and instrumentation have meant that
detailed studies of stellar formation, stellar evolution and chemical evolution 
have moved from the confines of our Galaxy and the Large and Small Magellanic
Clouds (LMC, SMC) to all the LG galaxies.  Imaging to faint limits in crowded
fields has been made possible with HST, and this has allowed us, with some 
difficulty, to reach the old main sequence turnoff in M31 and to measure RR
Lyraes throughout the LG.  
For distance scale work this has granted us the extra perspective 
resulting from the study of distance indicators 
in a variety of different environments. However interpreting
the observations is not
an easy task, as almost all galaxies contain multiple populations with
complex histories, and we now realize that interactions between many 
LG dwarf
galaxies and the two giant LG spirals are likely to have been a feature
throughout their lifetimes, as the present-time assimilation of the Sagittarius 
dSph
by our own Galaxy 
dramatically illustrates.  

Distances to galaxies in the LG are obviously needed as part of the study of
the galaxies themselves. Given the large dynamic range of astronomical
distances, which means that the distance scale is built up from overlapping
indicators starting with those we can
calibrate directly nearby, the LG galaxies play an essential role in the 
verification and extension of the distance scale.  In this short review we
will cover a selection of the recent work in the field; given the
huge amount of recent and on-going work on LG galaxies no attempt is made
to be complete and only work relating to the topic in hand will be addressed.  
For many of the lower luminosity galaxies 
our knowledge is still quite rudimentary,
albeit rapidly increasing due to the efforts by several groups. 
In section 2 we comment briefly on distance indicators relevant to the present
topic, and then in sections 3 through 6  discuss companions to our own Galaxy,
M31 and
its companions, luminous isolated galaxies, and finally faint isolated galaxies.
We conclude with a short summary. Note that a previous 
discussion of this topic is \cite{feast1}, and a convenient table listing LG 
galaxies and their distances from our Galaxy and the LG barycenter is found in \cite{vdb2}.
An extensive database of distances and other useful information
 is contained in \cite{ferra}, as part of the
Distance Scale Key Project.  The below discussion relies heavily on \cite{vdb1}, \cite{vdb2}
for details and evaluation of work prior to 2000.

Two other comments are in order.  Firstly, the nomenclature for LG galaxies is clearly
a mess, with the tradition of nameing newly discovered dSph's after the constellation
in which they are found, and only that, is nonsensical and a hinder to computer
searches at the very least.  This is clearly a matter that the International
Astronomical Union should take up.    The second comment refers to errors.  Unless
stated specifically to the contrary, here and elsewhere errors refer to the error
associated with the measurement of a distance, and do not include an estimate of the
error of the accuracy of the calibration of the distance indicator used.  For the latter,
systematic errors dominate; these are difficult to evaluate, and are almost always
underestimated.

\begin{table}
\caption{Our Galaxy and its companions}
\begin{center}
\renewcommand{\arraystretch}{1.4}
\setlength\tabcolsep{5pt}
\begin{tabular}{lllll}
\hline\noalign{\smallskip}
Name & Type$^{\mathrm a}$ &$M_V^{\mathrm a}$& Populations$^{\mathrm b}$
  \\
\noalign{\smallskip}
\hline
\noalign{\smallskip}
Galaxy & SbcI-II & -20.9  & all  \\
LMC & Ir III-IV & -18.5&  all \\
SMC & Ir IV-V & -17.1 &  all \\
Sagittarius  & dSph & -14: &  intermediate, old\\
Fornax & dSph & -13.1 &  (young), intermediate, old\\
Leo I & dSph & -11.9 & (young), intermediate, (old?)\\
Leo II & dSph & -10.1 & (intermediate), old \\
Sculptor & dSph & -9.8 & (young, with gas), intermediate, old \\
Sextans& dSph & -9.5 &  intermediate, old\\
Carina & dSph & -9.4 & (young), intermediate, old \\
U. Minor& dSph & -8.9 & (intermediate?), old \\
Draco & dSph & -8.6 &  old \\
\hline
\end{tabular}
\end{center}
$^{\mathrm a}$ From \cite{vdb1}, \cite{vdb2} \\
$^{\mathrm b}$ Minority populations are bracketed.
\label{Tab1a}
\end{table}

\begin{table}
\caption{M31 and its companions}
\begin{center}
\renewcommand{\arraystretch}{1.4}
\setlength\tabcolsep{5pt}
\begin{tabular}{lllll}
\hline\noalign{\smallskip}
Name & Type$^{\mathrm a}$ &$M_V^{\mathrm a}$& Populations$^{\mathrm b}$
  \\
\noalign{\smallskip}
\hline
\noalign{\smallskip}
M31  & SbI-II & -21.2 & all \\
M32 & E2 & -16.5 & (intermediate), mostly old \\
NGC 205 & Sph & -16.4 & (young), mostly intermediate, (old) \\
And I & dSph & -11.8 & mostly old \\
And II & dSph & -11.8 & intermediate, old \\
And III & dSph & -10.2 & intermediate, (old) \\
And V & dSph & -9.1 & old \\
And VI & dSph & -11.3 & mostly old \\
And VII & dSph & -12.0 & mostly old? \\
NGC 147 & Sph & -15.1 & (young \& intermediate), mostly old \\
NGC 185 & Sph & -15.6 & (young), intermediate, old \\
\hline
\end{tabular}
\end{center}
$^{\mathrm a}$ From \cite{vdb1}, \cite{vdb2} \\
$^{\mathrm b}$ Minority populations are bracketed.
\label{Tab1b}
\end{table}

\begin{table}
\caption{Brighter isolated LG galaxies}
\begin{center}
\renewcommand{\arraystretch}{1.4}
\setlength\tabcolsep{5pt}
\begin{tabular}{lllll}
\hline\noalign{\smallskip}
Name & Type$^{\mathrm a}$ &$M_V^{\mathrm a}$& Populations$^{\mathrm b}$
  \\
\noalign{\smallskip}
\hline
\noalign{\smallskip}
M33 & Sc II-III & -18.9 & all \\
IC 10 & Ir IV & -16.3 & all, no globular clusters \\
NGC 6822 & Ir IV-V & -16.0 & all \\
IC 1613 & Ir V & -15.3 & all, no globular clusters \\
WLM & Ir IV-V & -14.4 & all \\
\hline
\end{tabular}
\end{center}
$^{\mathrm a}$ From \cite{vdb1}, \cite{vdb2} \\
$^{\mathrm b}$ Minority populations are bracketed.
\label{Tab1c}
\end{table}

\begin{table}
\caption{Fainter isolated LG galaxies}
\begin{center}
\renewcommand{\arraystretch}{1.4}
\setlength\tabcolsep{5pt}
\begin{tabular}{lllll}
\hline\noalign{\smallskip}
Name & Type$^{\mathrm a}$ &$M_V^{\mathrm a}$& Populations$^{\mathrm b}$
  \\
\noalign{\smallskip}
\hline
\noalign{\smallskip}
Pegasus & Ir V & -12.3 &(young), intermediate, old \\
Sag DIG & Ir V & -12.0 & young, intermediate, (old?) \\
Leo A & Ir V & -11.5 &(young), intermediate, old \\
Aquarius & Ir V &-10.9 &  young, intermediate, (old?)\\
Pisces & Ir/Sph & -10.4 & young, intermediate, (old?)\\
Cetus & dSph & -10.1 & intermediate, old? \\
Phoenix & Ir/Sph & -9.8 & all\\
Tucana & dSph & -9.6 & old \\
\hline
\end{tabular}
\end{center}
$^{\mathrm a}$ From \cite{vdb1}, \cite{vdb2} \\
$^{\mathrm b}$ Minority populations are bracketed.
\label{Tab1d}
\end{table}

\section{Relevant Distance Scale Calibrators}

Most of the distance indicators discussed elsewhere in this volume (q.v.) are 
relevant for use within the LG, and only a few general comments will be made here.
 The more massive LG systems, with the exception
of M32 have had continuing, if in some cases spasmotic,
star formation over their whole lifetimes and thus all `population I' and
`population II' indicators can in principle be observed.  The lower mass galaxies are mostly
dominated by a mixture of intermediate and older populations, and thus indicators
such as the brightness of the Tip of the Red Giant Branch (TRGB) and RR Lyraes are very useful, 
although for the more distance LG galaxies the latter are difficult to measure,
even with HST.   The metal-poor, low-mass irregulars with recent star formation
contain ultra-short period Cepheids, and these have been advocated \cite{dolph} as
a useful indicator for these systems.  Perhaps most importantly, the diversity
of galaxies allows inter-comparison between distance indicators in a wide
variety of environments.  

In summary, primary indicators used to find distances to
LG galaxies include: Cepheids, Mira variables, RR Lyraes, RGB clump, Eclipsing
Binaries and TRGB.  Secondary distance indicators whose zeropoint relies wholly
or partially on distances to LG galaxies provided by the primary indicators
includes Planetary Nebulae Luminosity Function (PNLF), Supernovae, Surface Brightness
Fluctuations (SBF), Globular Cluster Luminosity Function (GCLF), Novae, and Blue
Supergiants.  The distinction is not always absolute, for instance 
TRGB when  calibrated by distances to Globular Clusters which themselves are
tied to Hipparcos parallaxes of subdwarfs is primary, but if it is calibrated from the
brightness of the Horizontal Branch (HB) and thus dependent on the adopted luminosities of RR Lyraes,
then it is secondary.
Depending on the degree of the reader's belief in the underlying theory, all the secondary
indicators could be considered primary, in principle.

\section{Companions of our Galaxy}

There are 11 known companions to our Galaxy, although the status of Leo I and
Leo II is uncertain.  Of these the Sagittarius dSph is in
collision with our Galaxy, and thus plays little part in distance scale studies.
Its mean distance is $(m-M)_0 = 17.36 \pm 0.2$ from Mira variables \cite{white} 
and $(m-M)_0 = 17.18 \pm 0.2 $ from RR Lyraes. Given the extended structure of
the Sagittarius dSph, such numbers are not particularly meaningful. 

The LMC
by contrast is pivotal in distance scale work, and will be discussed in some
detail here, and elsewhere in this volume \cite{feas}.
The major use of the LMC is as a sanity check - it includes most of the popular
distance indicators and is close enough so that they can be studied in great
detail, yet is far enough away so that to a  first approximation its
contents  can all be considered to be at the same distance from us.  Recent
reviews \cite{walk1},  see also \cite{feast2}, discuss the topic in great detail,
however progress has been rapid with improvements to the primary calibrators
that have resulted in improved consistency.  The comprehensive figure in \cite{bene} showing 
results ranging from $(m-M)_0 = 18.1$ to 18.8, although a good historical
summary, is more pessimistic than need be.   The smaller moduli mostly come from
early results based on using Hipparcos parallaxes for the locally common
RGB clump stars, without realization that both age and abundance each have
a dramatic effect on the absolute magnitude of the clump.  Modeling of these
effects \cite{gs}, \cite{sg} has provided quantitative understanding of the
evolution of clump stars, and has shown the advantage of observing in
 the infrared K-band which additionally
 greatly reduces the significance of reddening corrections compared to observing
in the visible.
New results for both LMC cluster \cite{sara} and field \cite{pg}, \cite{alv} 
all give LMC moduli near 18.5.

The LMC distance gap between the traditional indicators, Cepheids and RR
Lyraes, has also narrowed \cite{clem}, with the mean RR Lyrae modulus now
$18.44 \pm 0.05$, even with the traditionally short
value given by statistical parallaxes of galactic field RR Lyraes included.
The realization in recent years that the galactic halo contains star streams,
possibly remnants of accreted dwarf galaxies, makes less certain the
assumption of velocity homogeneity assumed in the statistical parallax method.
We will adopt, see \cite{clem}
\begin{equation}
<M_V(RR)> = 0.21([Fe/H]+1.5)+0.62
\end{equation}

For Cepheids, the remaining questions are well summarized elsewhere in this volume
\cite{fouq}, \cite{feas};
the characterization
of the effect of metallicity on the PL relation zeropoint still
defies solution, and is the most important unknown. Cepheids are well-understood both
observationally and theoretically, and with fundamental astrometric \cite{bene}
and interferometric \cite{nord}, \cite{lane} observations to add to the Hipparcos
parallax measurements \cite{fecat}, the likelihood of there being a significant
systematic error in the (metal-normal) PL zeropoint seems remote.

Eclipsing binaries are a promising technique, with the issues very clearly
set out by \cite{harr}, who gives distances for ten SMC binaries found by OGLE
\cite{piet}, solving the
technical difficulty of getting enough large telescope time to measure the
radial velocities by observing all the stars at once using the wide-field fiber
spectrograph 2DF on the Anglo-Australian telescope.
The three LMC systems have been recently (re)discussed, see \cite{groen},
\cite{fitz}, \cite{ribas}.

There are still some disquieting problems \cite{sala}, and there are
still some systematic differences between calibrators that we would like to
understand better.  However, the evidence seems strong for an `intermediate'
LMC modulus, and here (Table 5) we adopt $(m-M)_0 = 18.52$. It is noteworthy 
that for the recent determinations 
by a variety of methods the error bars overlap, this gives confidence that there
are not undiscovered systematic errors, and so it seems not too unrealistic to
evaluate the overall accuracy of the above mean modulus as $\pm 0.1$ mag,
corresponding to $\pm 5\%$ in the distance.  Many of the estimates for other
LG galaxies below are tied to the LMC at a modulus of 18.50; we have made no adjustments
for the slight difference with the Table 5 value.

Turning now to the SMC, this galaxy has received far less prominence in
comparison to  the LMC, mostly due to the considerable
extent of the SMC along the line of sight.  The degree
of this extent is controversial, see \cite{cald} for a 3-D model.
The SMC Cepheids show considerable dispersion
in the period-luminosity (PL) relation, but there is little room from the small dispersion in
the period-color relation to allow for a significant range in reddening or
possibly metallicity, thus it is difficult to explain the PL dispersion as
anything other than a depth effect.  Even a `mean' distance to the SMC 
derived from different distance indicators may not be comparable if there
are differences in the spatial distribution of SMC stars as a function of age.
Despite this cautionary note, the SMC has mean metallicity substantially
lower than the LMC \cite{luck} and thus it is of use for investigating the
effects of metallicity on distance indicators \cite{feas}.
Earlier work, as summarized by \cite{walk1} gives $0.42 \pm 0.05$ for the
difference between the LMC and SMC moduli, a result largely based on
the Cepheids, thus $(m-M)_0 = 18.94$ for the LMC at 18.52.

The remaining galaxies in this group are all of type dSph, and with the 
exception of Fornax and Sagittarius are amongst the lower luminosity
examples of this type, which is likely a selection effect \cite{mat}.  
With their significant old populations, these galaxies all contain many
RR Lyraes.  We give some updates to the distance estimates tabulated
in \cite{vdb1}, \cite{vdb2}.  For Sculptor, using OGLE photometry\cite{kal} and assuming
mean $[Fe/H] = -1.9$ for the RR Lyraes, $(m-M)_0 = 19.59$, while restricting the
sample to just the double-mode
RRd stars, \cite{kov} finds $(m-M)_0 = 19.71$.

Photometry for 515 RR Lyraes in Fornax has recently been published \cite{ber}
who find $<V_0> = 21.27 \pm 0.10$, with $[Fe/H] = -1.6 \pm 0.2$, $(m-M)_0 = 20.67$.
This is in good agreement with their earlier work \cite{ber2} which gives a TRGB 
distance of 20.68 mag.

The most recent RR Lyrae photometry for the Carina dSph is by \cite{dall}.  With
$<V_0> = 20.68$ and assuming a  mean $[Fe/H] = -1.7$, $(m-M)_0 = 20.06 \pm 0.12$.
This value is in excellent agreement with earlier work \cite{vdb1}.

The distance to the Sextans dSph is given \cite{mat2} as $(m-M)_0 = 19.67 \pm 0.15$,
however there are uncertainties in the metallicity which could change this 
value. These authors also discovered
an intermediate age population as evinced by six anomalous Cepheids, and 
\cite{bel} further discuss the multiple populations and their metallicities.

The Draco and Ursa Minor dSphs have recently been compared \cite{bel2}, with
respective distances from the horizontal branch magnitude of $(m-M)_0 = 19.84 \pm 0.14$ and
$19.41 \pm 0.12$ being derived.  These distances are in good agreement with those
found by the TRGB method.

Leo I and Leo II are considerably more distant than the above, and despite
morphological similarities have strikingly different star formation histories 
\cite{gal2}.  The best distances to Leo I appear to be those measured using
the TRGB method \cite{lee3}, $(m-M)_0 = 22.16 \pm 0.08$, and from RR Lyraes
by \cite{held}, $(m-M)_0 = 22.04 \pm 0.14$.   Similar data are available
for Leo II, where \cite{vdb1} evaluates the
distance as $(m-M)_0 = 21.60 \pm 0.15$.  For Leo II, the discovery of copious
numbers of RR Lyrae variables \cite{sieg} will likely yield an improved distance.

\begin{table}
\caption{Distance Modulus Measurements for the LMC}
\begin{center}
\renewcommand{\arraystretch}{1.4}
\setlength\tabcolsep{5pt}
\begin{tabular}{lll}
\hline\noalign{\smallskip}
Indicator  & Value & Reference \\ 
\noalign{\smallskip}
\hline
\noalign{\smallskip}
Cepheids &    $18.55\pm0.06$ & \cite{fouq}, \cite{feas}        \\
RR Lyraes &   $18.44\pm0.05$ & \cite{clem}        \\
RG Clump &     $18.49\pm0.06$ & \cite{sara}, \cite{pg}, \cite {alv}  \\
TRGB       &   $18.59\pm0.09$  &\cite{cion}, \cite{rom}       \\
Eclipsing Bin.  &$18.46\pm0.1$ & \cite{groen}, \cite{fitz}, \cite{ribas}         \\
Miras       &  $18.59\pm0.2$ & \cite{white}         \\
SN 1987A & $ 18.55\pm0.17$ & \cite{walk1}           \\
\\
Mean & $ 18.52\pm0.10$ \\
\hline
\end{tabular}
\end{center}
\label{Tab2}
\end{table}

\section {M31 and its Companions}

M31 contains all the distance indicators mentioned above and, as well stated
by \cite{clem2} {\it An SbI-II giant spiral galaxy provides a much more appropriate
local counterpart to the Distance Scale Key Project galaxies than does the LMC...    
M31 is also an important calibrator for the PNLF zeropoint, and also for the
Globular Cluster Luminosity Function (GCLF) method, applicable to massive galaxies 
with large GC populations.
Therefore, in any respect except for ease of observations, M31 is a much more 
important cornerstone for the distance scale than the LMC.}  
To which might be
added the difficulties include both the variable (internal) reddening, and the
large angular extent on the sky, the latter now being addressed by the latest
generation of wide-field imagers and multi-object spectrometers.

The distance to M31 has long been established using Cepheids, with a much-quoted
result \cite{wendy}, referenced to the LMC at an assumed distance modulus of 18.50
and reddening E(B-V) = 0.10, of $(m-M)_0 = 24.44 \pm 0.10$.  From HST photometry of
M31 clusters, \cite{rich} found $V_0 (HB) = 25.06$ at $[Fe/H] = -1.5$, then with
$M_V(RR) = 0.62$, $(m-M)_0 = 24.44$, while from isochrone fits to the RGB, 
\cite{holl} found $(m-M)_0 = 24.47 \pm 0.07$.  All these results are in remarkably
good agreement.  A major effort that will improve the amount of data available for
M31 Cepheids and Eclipsing Binaries is the DIRECT Project \cite{macri} which has the aim
of measuring the distance to M31 in one-step via the Baade-Wesselink method for Cepheids
and by discovering and measuring a significant number of eclipsing binaries.

Using HST, \cite{clem2} have shown that it is possible to measure M31 cluster RR
Lyraes, but the observational task is less formidable for field RR Lyraes in the
companion galaxies to M31.  For instance \cite{prit} give HST lightcurves for 111
RR Lyraes in And VI, and derive intensity-mean $<V>_0 = 25.10 \pm 0.05$, with
$[Fe/H] = -1.58 \pm 0.20$  \cite{ajd}, and the RR
Lyrae magnitude-metallicity relation above, $(m-M)_0 = 24.50 \pm 0.06$.  The And VI
distance from the TRGB method, is
$(m-M) = 24.45 \pm 0.10$\cite{ajd}. Systematic HST photometry of other dSph companions
to M31 are yielding distances via the magnitude of the horizontal branch or mean magnitudes of the
RR Lyraes.  For And II, \cite{dac1} measure $(m-M)_0 = 24.17 \pm 0.06$, while for
And III they find \cite{dac2} $(m-M)_0 = 24.38 \pm 0.06$.  Clearly, with accurate
distances relative to M31 the true spatial distribution of the M31 dSph companions
can be mapped; this requires accurate photometry and a knowledge of
the metallicity.  

M32 is the closest companion to M31, it is a dwarf elliptical,
with clear indications of interactions and likely tidal stripping by M31 \cite{vdb1}.
It is an important site for stellar population studies, until the recent discovery
\cite{dav} of luminous AGB stars it was argued that M32 contained only an old population.
The distance to M32 is usually assumed to be the same as for M31 \cite{vdb1}.

NGC 205 is also a close companion of M31, distance estimates are well summarized 
by \cite{vdb1}, with for example a TRGB distance of 24.54 \cite{sala}.  HST CMDs for
NGC 205 clusters are discussed in a preliminary report by \cite{geis2}.

NGC 147 and 185 lie close together on the sky and the evidence is strong that they
are bound to each other \cite{vdb1}, less certain is whether they are bound
to M31 \cite{vdb4}.  Early distance measurements, including those via RR Lyraes, are summarized by
\cite{vdb1}.  The TRGB estimate for NGC 147 by \cite{sala2} is 24.27, they also give 
24.12 for NGC 185, with an independent TRGB estimate \cite{mart} of $23.95 \pm 0.10$, 
Both galaxies therefore are slightly closer to us than M31, and as pointed out by
\cite{vdb1}, lie close to the LG barycenter.

\section {Luminous Isolated LG Galaxies}

The spiral galaxy M33 is the third most luminous galaxy in the LG, although it is only 
slightly brighter than the LMC \cite{vdb1}.  Recent distance measurements have shown 
considerable dispersion, although it has been suggested \cite{jacoby} that they may all
be reconciled by reasonable adjustments of the reddening, and it will be interesting to
see whether or not that is indeed the case.  They also suggest that to circumvent the
reddening problem for Cepheids, a technique of determining the periods using optical 
photometry, followed by a single-epoch infrared K-band observation, should be used.  As the
phasing is known, the K-band observation need not be taken at random phase but can instead
be chosen to correspond to phases near mean light, since although the
K band amplitudes of Cepheids are small, they are not negligible.  Using periods from the
DIRECT Project \cite{macri} together with single-epoch HST I-band observations, \cite{lee}
find for 21 Cepheids $(m-M)_0 = 24.52 \pm 0.14 (stat) \pm 0.13 (sys)$ assuming $E(B-V) = 0.20$
for M33 and based on an LMC distance of 18.50 mag. and $E(B-V)_0 = 0.10$.  The Key Project
Cepheid distance, for 11 stars, is very similar at $24.56 \pm 0.10$.  Using the same HST
data set, \cite{kim} found a rather larger distance from RGB stars in multiple fields, 
$24.81 \pm 0.04 (stat) \pm 0.13 (sys)$  from the TRGB and $24.90 \pm 0.04 (stat) \pm 0.05 (sys)$
from the RGB clump.   Photometry of M33 halo clusters \cite{sara2} gives very similar values,
from the horizontal branch magnitude in two clusters $(m-M)_0 = 24.84 \pm 0.16$, while from 
the position of the RGB clump in 7 clusters $(m-M)_0 = 24.81 \pm 0.24$.  

There are four other relatively luminous isolated galaxies, all are Irregulars of type IV or
V.  Due to its very low galactic latitude and consequent high foreground reddening, IC 10
is difficult to study.  Reddening estimates in the literature range over a very wide value,
and to complicate matters the internal reddening seems highly variable, perhaps not
surprising given the high star formation rate.  From V and I observations of Cepheids \cite{sakai}
derive $(m-M)_0 = 24.1 \pm 0.2$, and $E(B-V) = 1.16 \pm 0.08$.  With this reddening, their
TRGB distance is $(m-M)_0 = 23.5 \pm 0.2$, but they regard this as a lower limit since there
is no reason to expect the halo of IC 10 to have reddening as high as the inner regions
where the Cepheids are located.  To force the TRGB distance to be the same as given by the
Cepheids implies that the IC 10 halo has reddening of $E(B-V) = 0.85$, which would then
be primarily the amount of galactic foreground reddening.  A new estimate of $(m-M)_0 = 
24.4$, with $E(B-V) = 0.77$, is given by \cite{zuck}, but with no details.  Clearly,
infrared measurements for the IC 10 Cepheids would be of value in reducing the distance 
error for this very interesting galaxy.

There do not appear to be any distance estimates for NGC 6822 more recent that those
evaluated by \cite{vdb1}, who derives $(m-M)_0 = 23.48 \pm 0.06$ from a weighted mean.  
Recently, \cite{ant} have found many more Cepheid variables in a survey, the reference
describes those in a single 3.77 x 3.77 arcmin field.

For IC 1613, \cite{vdb1} derives $(m-M)_0 = 24.3 \pm 0.1$. From the TRGB method, 
\cite{tik} find $(m-M)_0 = 24.53 \pm 0.10$, and also determine $[Fe/H] = -1.75$.  As a by-product of
the OGLE project \cite{udal3} measured 138 Cepheids in a central field, and compared to
distances from the RR Lyraes and the TRGB, and concluded that the distance is 
$(m-M)_0 = 24.20 \pm 0.02 (stat) \pm 0.07 (sys)$.  A similar study is that by \cite{dolph3},
who compare Cepheids, RR Lyraes, RGB clump stars. and TRGB using deep HST V and I 
photometry, to find $(m-M)_0 = 24.31 \pm 0.06$.  In later work, \cite{dol3}, \cite{dol4},
they examine the question of whether ultra-short period Cepheids (USPC's, Population I 
Cepheids with periods less than two days) are useful distance 
indicators, comparing the properties of such stars in the SMC, LMC, IC 1613, Leo A,
and Sextans A.
It has been long known that metal-poor systems with young populations contain
more USPC's than do more metal rich systems. 
They find that USPC's do indeed appear to be good distance indicators, with excellent
agreement between the USPC's and TRGB, RGB clump, longer period Cepheids, and RR
Lyraes for Sextans A, Leo A, IC 1613 and the SMC, but not for the LMC where such stars
appear to be 0.2 mag. too luminous.  In the LMC USPC's
are uncommon, and thus it is postulated that these stars are fundamentally different from
those in the more metal-poor systems. It is well-known that the light curve amplitudes are
much smaller for the LMC USPC's compared to those in the SMC, for example.
  
The WLM galaxy has a distance \cite{vdb1} of $(m-M)_0 = 24.83 \pm 0.1$ from several
Cepheid and TRGB estimates.  There are two recent measurements, \cite{rej} observed a field
with STIS on HST, reaching the level of the horizontal branch.  Assuming $[Fe/H] = -1.5$
and $M_V = 0.7$, they find $(m-M)_0 = 24.95 \pm 0.13$.  Reddening to WLM is low, they
adopt $E(V-I) = 0.03$.  A rather similar result is found by \cite{dol6}, who give
$(m-M)_0 = 24.88 \pm 0.09$ from HST WFC2 photometry.  

In conclusion, the luminous, isolated galaxies in the LG provide a wealth of information
relevant to the distance scale.  They are relatively rich, so that they contain good-sized
samples allowing statistically significant comparisons to be made, and environs sufficiently
different one to the other that metallicity and age effects can be investigated in
depth.  Such work is on-going.  Distances are in relatively good agreement for the galaxies
with low reddening, objects like IC 10 are clearly much easier to study in the infrared.

\section {Faint Isolated LG Galaxies}

This category consists of the faint dwarf irregulars: Pegasus, Aquarius,
Sag DIG and Leo A, together with the fainter `transition' objects Pisces and Phoenix,
plus two dwarf spheroidals: Tucana and Cetus.
For the dwarf irregulars, by definition, star formation has occurred at some level up to
the present time, however the occurrence of rare stages of star formation depends critically
on the star formation rate at any given time.  Even for more luminous galaxies this effect
is well-seen, an example is the lack of long-period
Cepheids in WLM compared to the situation in the rather similar galaxy Sextans A. 

The Pegasus dwarf irregular galaxy (DDO 216) appears to have had little attention
since the summary by \cite{vdb1}, who points out that differences in the reddening adopted
between the several studies he quotes means that the distance is not well determined,
and he adopts $(m-M)_0 = 24.4 \pm 0.25$.  Depending on the true distance, Pegasus may
possibly be a distant member of the M31 sub-group.

The most recent distance to the Aquarius dwarf irregular galaxy (DDO 210) is that of
\cite{lee8}, who from the TRGB method finds $(m-M)_0 = 24.9 \pm 0.1$
 
The Sagittarius Dwarf Irregular galaxy (Sag DIG) has a distance from the TRGB method
by \cite{lee4} of $(m-M)_0 = 25.36 \pm 0.10$, and as such it is the outermost galaxy in the
LG according to \cite{vdb2}.

Leo A has been studied recently by \cite{sch}, who found a distance by the TRGB method of
$(m-M)_0 = 24.5 \pm 0.2$, and by \cite{dol7}, who from HST observations measured the brightness
of the RR Lyraes, to find $(m-M)_0 = 24.51 \pm 0.07$.  

Pisces, also widely refered to as LGS 3, has been observed by \cite{mill}, who find from 
the TRGB, the brightness of the clump RGB stars, and the level of the horizontal branch,
that $(m-M)_0 = 23.96 \pm 0.07$.  Classified as a transition object, there is still active
star formation in a small area approximately 60 pc in diameter near the center of the galaxy.

The central regions of Phoenix were studied using HST by \cite{holt}, who measured the
level of the horizontal branch at $V (HB) = 23.9 \pm 0.1$ which using the calibration of equation (1)
corresponds to $(m-M)_0 = 23.3 \pm 0.1$.  Their TRGB distance is somewhat shorter, $(m-M)_0 =
23.11$, very similar to earlier results \cite{held2}, \cite{mart2}.

The Tucana dSph is one of the most isolated galaxies in the LG, TRGB distances 
\cite{cast}, \cite{sav2} average to $(m-M)_0 = 24.76 \pm 0.15$.  HST imaging of this
galaxy \cite{lav} has never been published in detail, the CMD appears to show a single, old
population.

Finally, the Cetus dSph galaxy was recently discovered \cite{whit} and the first stellar
populations study, from HST observations, has just appeared \cite{sara2}.  From the
TRGB method, $(m-M)_0 = 24.46 \pm 0.14$, for $E(B-V) = 0.03$, identical to the ground-based
distance found by \cite{whit} using the same method.

\section {Summary}

The LG is a very important place, where we can study galaxies in detail and thus extrapolate 
our findings to the Universe at large. and it is where we set up and verify the distance 
scale ladder.  With the development of large format imaging mosaics, and the advent of very
large telescopes with powerful spectrographs, together with the unique capabilities of HST,
there has been an explosion in the amount of high quality data available for LG galaxies,
while in parallel there has been substantial progress on the theoretical understanding for
most of the popular standard candles, and substantial improvements in
their calibrations.  Specifically, it appears that the `long-short' 
problem for the distance to the LMC has largely vanished.  Distances from the reliable 
indicators are now within one signa of each other, and although it is clear there are still
systematic differences, they have shrunk, and the LMC modulus of $(m-M)_0 =
18.52 \pm 0.10$ seems reasonably secure.  Reduction in the size of the error, and improvement in 
the agreement between
distance indicators, will be aided by comparisons made in a variety of environments, and
here the LG galaxies are of key value.   The wealth of new work reported above will be invaluable
in this respect.

%

\end{document}